%% file: manuscript.tex
\RequirePackage{lineno}
\newif\ifpreprint%
\preprinttrue%\preprintfalse
\ifpreprint%
	\documentclass[twocolumn,letterpaper,english,aps,prb,amsmath,amssymb,superscriptaddress]{revtex4}
\else%
	\documentclass[twocolumn,letterpaper,english,aps,prb,floatfix,amsmath,amssymb,linenumbers,superscriptaddress]{revtex4}
\fi%

\usepackage{xr}
\usepackage{array}
\usepackage{braket}
\usepackage{bm}
\usepackage{color}
\usepackage{dsfont}
\usepackage{float}
\usepackage{graphicx}
\usepackage{units}

\ifx\pdftexversion\undefined%
\usepackage[dvips]{hyperref}
\else%
\usepackage{hyperref}
\fi%
\hypersetup{
  colorlinks = false,
  hypertexnames=false
}
\newcommand{\ssm}{\scriptscriptstyle\rm}

\renewcommand{\theta}{\vartheta}
\renewcommand{\phi}{\varphi}

\begin{document}
	
%\ifpreprint%
	%\linenumbers%
%\fi%

\title{\input{title}\unskip}

\date{\today}

\author{Doruk Efe G\"okmen}
	\affiliation{Institute for Theoretical Physics, ETH Zurich, 8093 Zurich, Switzerland}
\author{Zohar Ringel}
	\affiliation{Racah Institute of Physics, The Hebrew University of Jerusalem, Jerusalem 9190401, Israel}
\author{Sebastian D. Huber}
	\affiliation{Institute for Theoretical Physics, ETH Zurich, 8093 Zurich, Switzerland}
\author{Maciej Koch-Janusz}
	\affiliation{Institute for Theoretical Physics, ETH Zurich, 8093 Zurich, Switzerland}
	\affiliation{Department of Physics, University of Zurich, 8057 Zurich, Switzerland}
	\affiliation{James Franck Institute, The University of Chicago, Chicago, Illinois 60637, USA}

%TC:ignore 
%%% Abstract
\begin{abstract}
	Identifying the relevant degrees of freedom in a complex physical system is a key stage in developing effective theories in and out of equilibrium. The celebrated renormalization group provides a framework for this, but its practical execution in unfamiliar systems is fraught with ad hoc choices, whereas machine learning approaches, though promising, lack formal interpretability. Here we present an algorithm employing state-of-art results in machine-learning-based estimation of information-theoretic quantities, overcoming these challenges, and use this advance to develop a new paradigm in identifying the most relevant operators describing properties of the system. We demonstrate this on an interacting model, where the emergent degrees of freedom are qualitatively different from the microscopic constituents. Our results push the boundary of formally interpretable applications of machine learning, conceptually paving the way towards automated theory building.
\end{abstract}

\maketitle
%TC:endignore 

%%

Fundamental physical theories, in a reductionist spirit, are often formulated at the smallest scales, describing the interactions of elementary constituents. Nonetheless, the experimentally accessible features typically arise from their collective behaviour. Indeed, there exist profound examples of \textit{effective} theories, \textit{e.g.}~classical hydrodynamics and thermodynamics, consistently describing complex phenomena in terms of a few macroscopic variables, without making 
any reference to individual particles.

Bridging this scale gap to \emph{derive} the emergent macroscopic properties  from microscopic models is a perpetual challenge. The renormalization group (RG),\cite{PhysicsPhysiqueFizika.2.263,Wilson1974,Wilson1975,Fisher1998} provides a powerful framework for this, associating physical theories at different length scales by iteratively coarse-graining configurations of local degrees of freedom (DOFs). The induced RG transformation acts as a telescope in the space of models, generating the RG flow, whose structure around the fixed point eventually reveals the relevant DOFs. They are the scaling operators, which determine the correlations, and thus the physical properties, at large scales.

In practice, executing this program in real-space RG is very difficult. The accuracy of the procedure is improved by optimizing the coarse-graining to retain the highest real-space mutual information (RSMI),\cite{Koch-Janusz2018,optimalRSMI} quantifying correlations to distant parts of the system. However, this still misses a crucial insight: any long-range information is due to the scaling operators and thus its optimal compression not only can serve as a better RG transformation, but should allow to extract all the operators themselves, without ever explicitly executing the RG flow. This was recently proven in part: in critical systems, the formal solutions to the RSMI compression problem are determined by the most relevant operators,\cite{Gordon2020} \emph{in theory} allowing to access them directly. Unfortunately, solving this mathematical problem is notoriously hard in a general setting.\cite{poole2019variational}

\begin{figure}[h!]
	\includegraphics{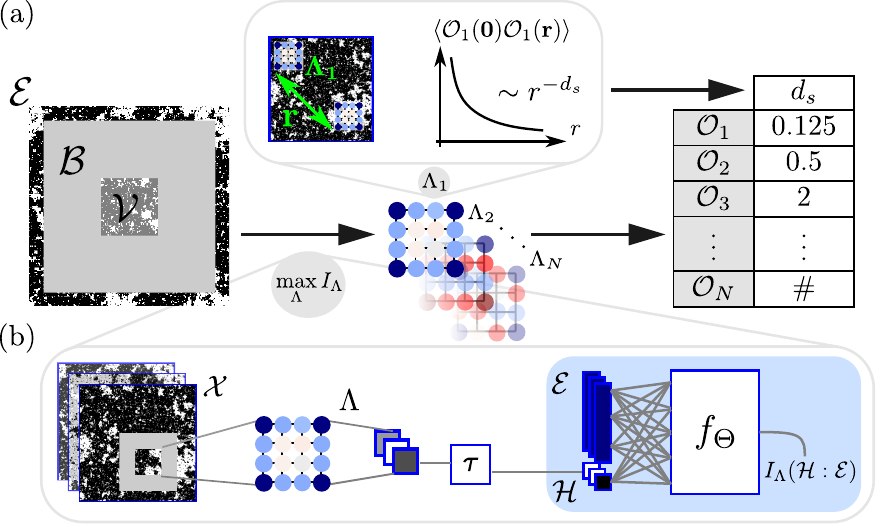}
	\input{fig1caption}
	\label{fig:concept}
\end{figure}

Here, using state-of-art machine learning results in estimating mutual information,\cite{poole2019variational} we overcome this challenge to develop a highly efficient algorithm extracting relevant operators of the theory from real-space configurations. In contrast to standard approaches no RG maps are iterated: scaling operators are \emph{not} constructed from the RG flow, but instead using their definition as dominant contributions to RSMI, \emph{in a single step}. The RSMI neural estimator (RSMI-NE) returns them parametrized as neural networks, which can be assigned scaling dimensions and used in computations (see Fig.~\ref{fig:concept}.a). Moreover, we empirically demonstrate the power of the method across the whole phase diagram, also away from criticality.

In particular, the algorithm can, unsupervised, construct order parameters, locate phase transitions, and identify spatial correlations and symmetries for complex and large dimensional real-space data. Our findings, elevating the coarse-graining transformations to formal operators, give a new paradigm in investigating statistical systems, and a numerical toolbox to do so.

An often raised criticism of the use of machine learning in physics is the lack of interpretability of the results.\cite{doi:10.1098/rsta.2016.0153} Particularly, the extent to which architecture- and training-dependent conclusions from machine learning relate to formal concepts in physical theories is unclear. RSMI-NE overcomes this challenge: its outputs are explicitly identified with the scaling operators\cite{Gordon2020} on the lattice.  Thus, in contrast to generic data-driven approaches, RSMI-NE executes a \textit{physical} principle using machine learning tools to produce theoretically interpretable results.

Below we give an overview of the general RSMI setup, introducing the probabilistic language of the coarse-graining optimisation. We then present the RSMI-NE algorithm, and the theoretical and numerical results in machine learning underlying its efficiency. We validate its capabilities on an interacting model, whose non-trivial RG flow was a subject of a detailed theoretical analysis.\cite{PhysRevLett.94.235702,PhysRevE.74.041124} We investigate the physical data contained in the ensemble of RSMI filters. We conclude with a discussion of further applications, most notably to non-equilibrium problems.

Consider a system of classical DOFs in any dimension denoted by a collective random variable $\mathcal{X}$, whose physics is specified by a probability measure $p(x)$, either Gibbsian dictated by the energy of the realization $x$ of $\mathcal X$, or a generic non-equilibrium distribution. A coarse-graining (CG) rule $\mathcal{X}\to \mathcal{X}'$ is defined as a conditional distribution $p_\Lambda(x'|x)$,  determined by a set of parameters $\Lambda$ to be optimised. The coarse-graining is typically carried out on disjoint spatial blocks $\mathcal{V}_i \subset \mathcal{X}$, and it factorises:  $p(x'|x)=\prod_i p_{\Lambda_i}(h_i|v_i)$, such that $\mathcal{X}=\bigcup_i \mathcal{V}_i$ and $\mathcal{X}'=\bigcup_i \mathcal{H}_i$, with $p_{\Lambda_i}(h_i|v_i)$ the CG rule applied to block $i$. In translation invariant systems a fixed $\Lambda_i \equiv \Lambda$ suffices; with disorder each block can be individually optimised.

The RSMI principle identifies CG rules extracting the most relevant long-range features as the ones retaining the most information shared by a block $\mathcal{V} \subset \mathcal{X}$ to be coarse-grained, and its distant environment $\mathcal{E}$,\cite{Koch-Janusz2018,optimalRSMI} \emph{i.e.}~those that optimally \emph{compress} this information. The environment is separated from $\mathcal{V}$ by a shell of non-zero thickness constituting the buffer $\mathcal{B}$, and forms the remainder of the system (see Fig.~\ref{fig:concept}.a). The ``shared information" between the random variables $\mathcal{H}$ and $\mathcal{E}$ is given by the mutual information:
\begin{equation}\label{eq:RSMI_def}
	I_{\Lambda}(\mathcal{H}:\mathcal{E}) = \sum_{h,e}p_\Lambda(e,h) \log\left(\frac{p_\Lambda(e,h)}{p_\Lambda(h)p(e)}\right),
\end{equation}
where $p_\Lambda(e,h)$ and $p(h)$ are the marginal probability distributions of $p_\Lambda(h,x)=p_\Lambda(h|v)p(x)$ obtained by summing over the DOFs in $\{\mathcal V$, $\mathcal B\}$ and $\{\mathcal V, \mathcal B, \mathcal E\}$, respectively. Finding such optimal coarse-graining requires thus maximizing $I_\Lambda$ as a function of parameters $\Lambda$.

The conceptual importance of the buffer $\mathcal{B}$ cannot be overstated: it sets the length-scale filtering out contributions of short-range correlations between $\mathcal{V}$ and $\mathcal{E}$. Increasing its thickness $L_\mathcal{B}$ corresponds to growing the RG scale, preserving only the long-range physics. With an arbitrary fixed CG rule this can only be achieved in RG by iterating the coarse-graining, with all the ensuing difficulties, particularly amplifying the errors in the formulation of the rule. In our approach \emph{the CG rules themselves contain long-range information}, and are obtained in a single shot, by solving the $I_\Lambda$ optimization problem directly at large $L_\mathcal{B}$, at different points in the phase diagram.

The optimization problem of Eq.~(\ref{eq:RSMI_def}) is, however, difficult, as estimating or maximizing mutual information is notoriously hard.\cite{poole2019variational} 
This was a major weakness of the RSMI proposal,\cite{Koch-Janusz2018} hindering numerical and theoretical progress.
We can now overcome this challenge. At the heart of our approach, encapsulated in the RSMI-NE algorithm, is a series of recent results combining mathematically rigorous variational bounds on mutual information \cite{doi:10.1002/cpa.3160360204,NIPS2003_2410,5605355} with deep learning.\cite{belghazi2018mine,poole2019variational} A family of \emph{differentiable} lower bounds to $I_\Lambda$ is introduced, parametrised by neural networks $f_{\Theta}$ (see Fig.~\ref{fig:concept}.b), which in the course of gradient descent training on the joint samples of $\mathcal{H}$ and $\mathcal{E}$ become accurate, and in the limit exact, estimators of $I_\Lambda$, see the Supplemental Material (SM).\footnote{See Supplemental Material for details about the dimer model, more technical details of the RSMI-NE algorithm and its dependence on length scales, where also Refs.~\onlinecite{oord2018representation, 899a65b4919f47c8a06d115df85dad11, Goodfellow-et-al-2016, NIPS2014_5449, gumbel1954, kingma2014adam} are included.} The transformation $p_\Lambda(h|v)$ feeding the coarse-grained variables into the estimator is also expressed by a neural network \emph{ansatz}. We use the following composite architecture (see Fig.~\ref{fig:concept}.b): 
\begin{equation}
h = \tau \circ (\Lambda \cdot v).
\end{equation}
Here $\Lambda$ become parameters of a convolutional neural network (CNN) applied to the configurations, and $\tau$ differentiably maps $\Lambda \cdot v$ into states of variable $h$ of pre-determined type (\emph{e.g.}~pseudo-binary spins). This last embedding step is both crucial,\cite{PhysRevX.10.031056} and algorithmically non-trivial.\cite{jang2016categorical} We emphasize that while the CNN choice is motivated by convenience, any sufficiently expressive \emph{ansatz} can be used.

\begin{figure*}[t]
	\includegraphics{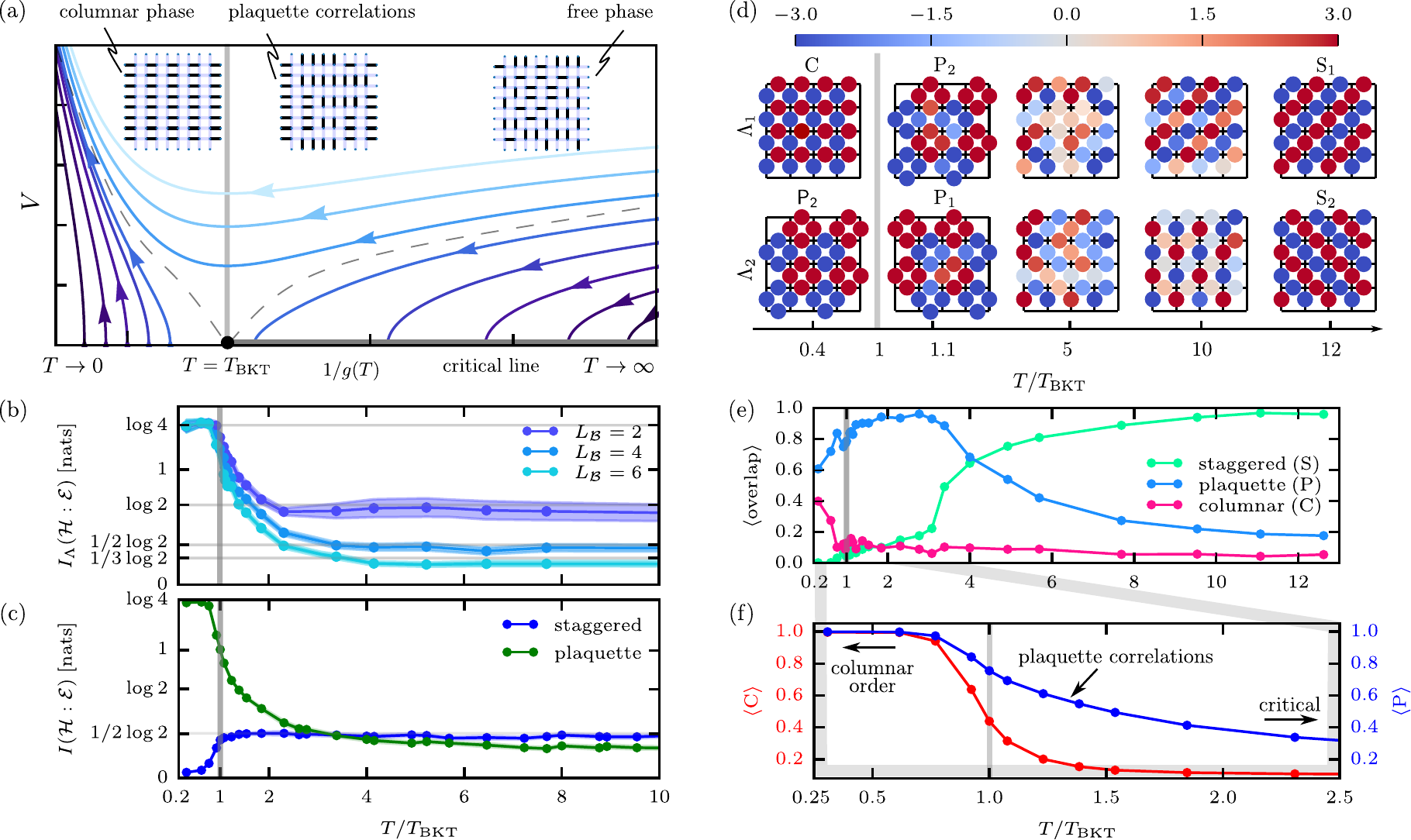}
	\input{fig2caption}
	\label{fig:main_dimer}
\end{figure*}

We have thus cast \emph{both} the CG rule and the lower-bound to the cost function it optimizes as differentiable neural networks. Next, we can chain them together (see Fig.~\ref{fig:concept}.b), and \emph{simultaneously} optimize via stochastic gradient descent, improving the RSMI estimator and the CG rule in each pass. Note, that it is this numerical breakthrough which enables the exploration of new theoretical ideas and renders the RSMI algorithm a promising new approach to tackle open challenges in complex domains.

We demonstrate this on the example of an interacting dimer model. This is an optimal test-bed for the illustration of our algorithm. First, a large class of classical statistical physics problems can be mapped to interacting dimer models.\cite{doi:10.1063/1.1703953,Blote_1982,Henley_1997,10.2307/2692028,Kenyon_2002,Cimasoni_2007,PhysRevLett.101.155702,PhysRevB.80.045112,PhysRevB.80.134413,PhysRevLett.122.080601} Moreover, aspects of the quantum dimer model\cite{PhysRevLett.61.2376,PhysRevB.69.224415} leave their footprint on the phase diagram.\cite{PhysRevLett.94.235702,PhysRevE.74.041124}
Second, in the dimer model, the relevant low-energy degrees of freedom are profoundly different form the microscopic building blocks of the theory and change qualitatively throughout the phase diagram. Hence, the algorithm is presented a non-trivial task.

The model is defined by the partition function $Z(T)=\sum_{\{C\}}\exp{(-E_C/T)}$ at a given temperature $T$ and
the configurations $C$ involve binary-valued microscopic degrees of freedom, dimers, that sit on the edges of the square lattice. They obey the constraint of exactly one dimer being connected to every vertex.
The energy $E_C=N_C(||)+N_C(=)$ counts plaquettes covered by parallel dimers favoured by the interaction, see Fig.~\ref{fig:main_dimer}.a. 

The essence of this system is in the interplay of aligning interaction energy and entropic effects due to the non-local cooperation of local dimer covering constraints. At low $T$ the former facilitates a long-range order (LRO) crystallizing the system into one of the four translation symmetry breaking \textit{columnar} states, see Fig.~\ref{fig:main_dimer}.a. With increasing $T$ the system undergoes a Berezinskii-Kosterlitz-Thouless (BKT) transition at $T_{\ssm BKT}= 0.65(1)$,\cite{PhysRevE.74.041124} entering a critical phase characterised by algebraic decay of correlations (also at $T\to \infty$) with exponents continuously changing with $T$. 
This is reflected in the effective continuum field theory, which, via the mapping of dimer configurations to height-field $\varphi(\mathbf{r})$\cite{fradkin_2013} (see also SM for the definition) is given by a sine-Gordon (SG) action:\cite{PhysRevE.74.041124}
\begin{equation}\label{eq:SG_dimer}
	S[\varphi(\mathbf{r})] = \int \mathrm{d}^2 \mathbf{r}\left[\frac{g(T)}{2}|\nabla \varphi(\mathbf{r})|^2 + V \cos\left( 4 \varphi(\mathbf{r})\right) \right].
\end{equation}
The potential $V$ locks $\varphi(\mathbf{r})$ into four values corresponding to the columnar states. The stiffness $g(T)$ controls fluctuations of $\varphi(\mathbf{r})$: large $g(T)$ favours ``flat" fields of high-entropy, low $g(T)$ allows large gradients corresponding to the staggered configurations, which are not suppressed in the algebraic phase.
The RG flow is shown in Fig.~\ref{fig:main_dimer}.a: the $T<T_{\ssm BKT}$ fixed point with finite $g(T)$ and $V\to\infty$ leaves energy minimisation as the sole relevant constraint; the \emph{line} of fixed points at $V=0$ at $T>T_{\ssm BKT}$ indicates that the energetic interactions are irrelevant and exponents vary with $T$. The flow reveals the physical nature of the algebraic correlations: $\nabla \varphi(\mathbf{r})$ obeys Gauss' law and so the fixed point theory is that of electrical fields.

To showcase RSMI-NE, we input Monte Carlo samples of the model across the whole temperature range to the algorithm. For concreteness, we restrict the coarse-grained variables $\mathcal{H}$ to a two-component binary vector $\{\pm 1,\pm 1\}$ (the optimal dimensionality can be found systematically\cite{rsmine}). Hence, we are looking for a two component vector of filters $\Lambda_1$,  $\Lambda_2$ determining how the visible region $\mathcal V$ is mapped onto $\mathcal H$. Optimizing the filters $\Lambda_1$,  $\Lambda_2$ for all $T$ separately gives a comprehensive picture of the long-wavelength physics, culminating in the construction of the relevant operators on the lattice, as we now show.

First, we find that already the curve $I_{\Lambda}(T)$, \emph{i.e.}~the amount of long-range information attained \emph{with the optimal} $\Lambda$, reveals the structure of the phase diagram (see Fig.~\ref{fig:main_dimer}.b). To wit, for $T<T_{\ssm BKT}$ its value is constant and equal to $\log 4$. The information shared between distant parts of the system in the ordered phase is precisely which of the four columnar states they are in. 

Phase transitions are reflected by non-analyticities in $I_{\Lambda}(T)$ (\emph{cf.} \onlinecite{Wilms_2011, PhysRevE.87.022128}). Moreover, the algebraic decay of $I_{\Lambda}(T)$ with the buffer size for $T>T_{\rm BKT}$ is indicative of a critical phase, see Fig.~\ref{fig:main_dimer}.b and Fig~6.c in Ref.~\onlinecite{rsmine}. This behaviour should also be contrasted with the exponential decay for the paramagnetic phase of 2D Ising model in Ref.~\onlinecite{rsmine}.

Going beyond the mutual information and examining the filters $\Lambda(T)$ yields further insight about spatial correlations. As conjectured, the optimal CG rules depend on the tuning parameters of the system. In the high- and low-temperature limits, three classes of filters emerge: independent optimizations return exclusively sets of $\Lambda_{1,2}$ that correspond to columnar and plaquette  at low temperatures, and staggered ones at high temperatures, see Fig.~\ref{fig:main_dimer}.d. We call these filters ``pristine'' as they reflect simple limiting cases.
They are orthogonal to each other and represent independent degrees of freedom. The filters for intermediate temperatures and their overlap with the pristine ones is shown in Figs.~\ref{fig:main_dimer}.d and e, respectively.
We first discuss in detail the individual filters $\Lambda_{1,2}$ in the different temperature regimes $T\to 0$, $T\sim T_{\ssm BKT}$ and $T\gg T_{\ssm BKT}$, and then explicitly match them with the RG-relevant operators of the continuum sine-Gordon theory.

The pristine plaquette and columnar filters at $T\to 0$ break translation or rotation symmetry, respectively. Any pair of $\Lambda_{1,2}$ drawn out of these filters is a \emph{bijection} between the four ordered columnar states and the four states $(\pm 1,\pm 1)$ taken by the compressed degrees of freedom $\mathcal H$. This degeneracy of plaquette and columnar filters is lifted when the rotation symmetry is restored: the pristine columnar filter is not found above $T_{\ssm BKT}$. Strikingly, its modulus acquires an expectation value for $T<T_{\ssm BKT}$ (see Fig.~\ref{fig:main_dimer}.f). This filter is thus an order parameter \emph{discovered} by RSMI-NE, and is in fact equal to the dimer symmetry breaking (DSB) order parameter identified in Ref.~\onlinecite{PhysRevLett.94.235702}.

The optimal CG rules around $T_{\ssm BKT}$ hold yet further insights. Particularly, the plaquette filters give rise to a \emph{putative} plaquette order parameter (see Fig.~\ref{fig:main_dimer}.g). The corresponding regime where it attains a non-vanishing value does not survive in the thermodynamic limit. However, the non-zero expectation value at finite system sizes (see Fig.~\ref{fig:main_dimer}.f) reveals the importance of such plaquette correlations, which are stabilised in the quantum dimer model (QDM).\cite{PhysRevLett.61.2376}
RSMI-NE indicates this without any prior insights about QDM, which inspired previous studies.\cite{PhysRevLett.94.235702,PhysRevE.74.041124}

Finally, the critical phase $T>T_{\ssm BKT}$ interpolates between pristine plaquette and staggered filters, due to the competition between the electric field operator and plaquette correlations in the finite system, as per Eq.~(\ref{eq:SG_dimer}). The value of RSMI attained with \emph{fixed} rules reflects this competition: the plaquette filter retains more information until well above $T_{\ssm BKT}$, where the staggered one takes over as plaquette correlations dwindle (see Fig.~\ref{fig:main_dimer}.c). The staggered filters \emph{are} the electric fields \emph{viz.}~they define coarse-grained variables $E_{1,2}(\mathbf{r}):=\tau \circ \Lambda_{1,2} \left(\mathcal{V}(\mathbf{r})\right)$, which precisely target the operator $\nabla \phi(\mathbf{r})$ (see SM).

\begin{table}
	\centering
	\caption{Pairs of filters drawn from the columnar and plaquette coarse-graining rules unambiguously label each of the four columnar ground states. The mapping is given by the sign of the scalar product of the filter (blue=-1, red=+1) with a dimer configuration (1 for occupied, 0 for unoccupied link). As the columnar configurations correspond to uniform height field, the electric charge operators $\mathcal{O}_{n=1,2}$, acting on the height-field $\varphi$ also serve as order parameters for the columnar phase and they directly correspond to the RSMI-optimal filters at low $T$.}
	\includegraphics[width=\linewidth]{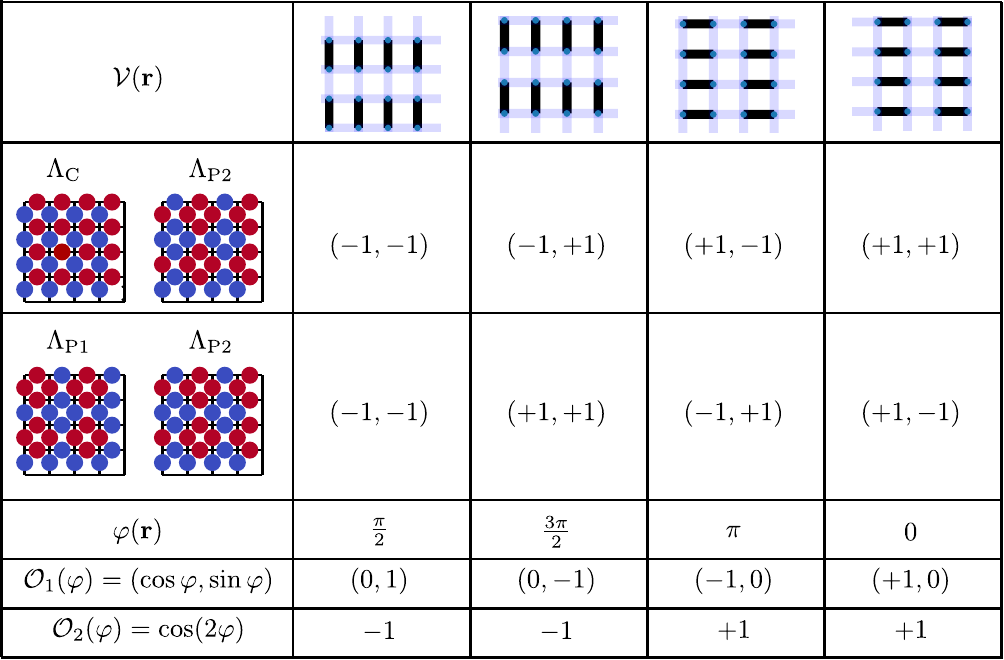}
	\label{fig:groundstatemaps}
\end{table}

The RSMI-NE finding the order parameters or the electric fields is no accident: the pristine $\Lambda$ filters define the relevant operators on the lattice. The considerable technical machinery behind this is the subject of Ref.~\onlinecite{Gordon2020}, here we show it using the field theory of the dimer model, also away from criticality. To wit, the  columnar and the DSB order parameters in Fig.~\ref{fig:main_dimer}.f correspond to the relevant electric charge operators $\mathcal{O}_{n}(\varphi)=(\cos(n\varphi), \sin(n\varphi))$,\cite{PhysRevB.76.134514} for $n=\pm 1$ and $n=\pm 2$, respectively. This is seen explicitly, using the height-field map in Table \ref{fig:groundstatemaps} as a dictionary:
\begin{align}
	\left(\Lambda_{\rm P1}, \Lambda_{\rm P2} \right)\circ \varphi &= \left(\cos(\varphi+3\pi/4), \sin(\varphi + 3\pi/4) \right), \\
	\Lambda_{\rm C}\circ \varphi &= \cos(2\varphi),
\end{align}
where on the left dimer configurations (on which the $\Lambda$ act) mapped to height field value $\phi$  are denoted by $\phi$ itself. 

Though competing correlations, especially in finite-size systems, may result in mixing of the pristine components, they can be identified by applying standard machine learning tools \emph{to the ensemble of filters}. Note that RSMI-NE is a stochastic algorithm, and through independent runs produces a distribution of optimal CG rules. 
Thus Fig.~\ref{fig:main_dimer}.d shows a sample of filters at each $T$, and in Fig.~\ref{fig:main_dimer}.e the overlap is averaged over the filter ensemble at each temperature. 
The distribution contains crucial information, \emph{e.g.}~the disappearance of the columnar filter above $T_{\ssm BKT}$ signals the lifting of the columnar/plaquette degeneracy (consistent with the scaling dimensions of $\mathcal{O}_{n}$, which go as $n^2$), 
and restoration of the rotation symmetry. 
More concretely, representations of the broken symmetries can be identified in the distribution, whereas at high-$T$ it can be used to retrieve even the \textit{emergent} $U(1)$ symmetry of the electrical field! See Ref.~\onlinecite{rsmine} for a more detailed discussion.

We thus managed to automatically sequence the operators of the theory, returning their lattice representations which are modular, reusable and may be formally labelled by their scaling dimensions. Indeed, evaluating a correlator of two neural networks parametrized by the plaquette filters, we fit a scaling dimension of $1.00037$  at $T\rightarrow\infty$ (see SM), in excellent agreement with $1.0$ predicted for $\mathcal{O}_{1}$.\cite{PhysRevB.76.134514} This raises the remarkable prospect of building a complete effective theory from raw data using machine learning. 

Though the discussion centred around an equilibrium example in two dimensions, our procedure works in any dimension, can be adapted to disorder,\cite{PhysRevX.10.031056} and does not require the existence of a Hamiltonian, as it only uses probability distributions. While a formal understanding of this approach for non-equilibrium distributions, extending the results of Ref.~\onlinecite{Gordon2020}, is missing, in the companion paper\cite{rsmine} we validated  the concept on the example of lattice model with aggregation and chipping,\cite{PhysRevE.63.036114} for which RSMI-NE locates precisely the non-equilibrium phase transition. We believe complex systems, such as realized in \emph{e.g.}~active matter\cite{PhysRevE.58.4828,Bialek4786} or atmospheric phenomena,\cite{Peters2006} to be a natural arena where information-theoretic methods can be applied,\cite{Dewar_2003} and our conceptual and numerical advancements may provide new theoretical insights (see also Ref.~\onlinecite{Nir30234}). The understanding of challenging higher dimensional interacting and quasiperiodic statistical systems \cite{PhysRevB.82.085114,PhysRevB.86.214414,quasiperiodic_spin_chains,PhysRevX.10.011005} may also benefit from this new method. 

%TC:ignore 

\bigskip
\noindent
{\bf Code availability} Source code for the RSMI-NE is available online at \url{https://github.com/RSMI-NE/RSMI-NE}.

\smallskip
\noindent
{\bf Acknowledgements} M.K.-J. is grateful to F. Alet for his comments on the physics of the interacting dimer model. D.E.G., S.D.H., and M.K.-J. gratefully acknowledge financial support from the Swiss National Science Foundation and the NCCR QSIT, and the European Research Council under the Grant Agreement No.~771503 (TopMechMat), as well as from European Union's Horizon 2020 programme under Marie Sklodowska-Curie Grant Agreement No.~896004 (COMPLEX ML). Z.R. acknowledges support from ISF grant 2250/19. Some of the computations were performed using the Leonhard cluster at ETH Zurich. This work was supported by a grant from the Swiss National Supercomputing Centre (CSCS) under project ID eth5b.

%\smallskip
%\noindent
%{\bf Author contributions} ...

%\smallskip
%\noindent
%{\bf Supplementary Information} for this paper is available online.

\ifpreprint%
	\clearpage

	\appendix
	\begin{widetext}
		\begin{center}
			{\normalsize \bf Supplemental Material: \input{title}\unskip}
		\end{center}
		
		\input{supp_text_prl}
	\end{widetext}
	\clearpage
\else%	

\fi%

\bibliographystyle{naturemag}
\bibliography{ref}

%TC:endignore 

\end{document}

%% file: title.tex
Statistical physics through the lens of real-space mutual information

%% file: fig1caption.tex
\caption{
{\bf Extracting the relevant operators with RSMI NE.} (a) The most relevant operators are \emph{learnt} as the optimal compressions of long-range information $I(\mathcal{H}:\mathcal{E})$ at each point in the phase diagram. The learnt maps can be associated to the physical operators by computing the correlators and extracting the scaling dimensions.
%Long-range physics is thus extracted at the outset of RG. 
(b) The architecture of RSMI-NE: the relevant operators are extracted via the transformations $\Lambda$ and discretizing step $\tau$. The long-range information which $\Lambda$ maximize is estimated by $f_\Theta$, all of which are parametrized by neural networks and co-trained together.
} 

%% file: fig2caption.tex
\caption{
{\bf RSMI analysis of the interacting dimer model.} (a) RG flow of the model (see Eq.\ref{eq:SG_dimer}) and representative configurations (top panel). (b) Total RSMI extracted with the optimal filters as a function of $T$ and its scaling with the buffer size. (c) Information extracted by the pristine staggered and plaquette filters at different T. (d) Samples of optimal filters obtained with RSMI-NE for different $T$ [columnar (C), plaquette (P1, P2) and staggered (S1, S2)]. (e) The average overlap of the optimal filters at $T$ with the pristine components. (f) The dimer symmetry breaking and plaquette order parameters extracted using the low-$T$ pristine filters.
} 

%% file: supp_text_prl.tex
\section{RSMI optimization}\label{section:methods}

Here we provide a description of the algorithmic components of the RSMI-NE, referencing the necessary definitions and results from machine learning and statistics. The estimation of of real-space mutual information (RSMI) is three to four orders of magnitude faster than in Ref.~\onlinecite{Koch-Janusz2018}; allowing to explore large systems (we tested up to the size of 256x256), and the limit of large buffer, while showing superior convergence and stability. The typical run-times are \emph{ca.} 30 seconds. This dramatic improvement in performance is due to very recent advances in estimating information theoretic quantities,\cite{belghazi2018mine,poole2019variational} and it, in turn, enables development of an entirely new approach to real-space renormalization based on RSMI, in particular  enabling the extraction of relevant operators (see main text, and section \ref{section:dimer_further}).

\subsection{Some properties of mutual information}
The Shannon \textit{mutual information} (MI) of two random variables $X$ and $Y$ quantifies the amount of knowledge we gain about one of them, when observing the other. Formally it is defined as a difference of entropies:
\begin{equation}
	I(X:Y):=H(X)-H(X|Y) = H(X) + H(Y) - H(X,Y).
\end{equation}
Typically RSMI can take values at most on the order of a few units of information when coarse-graining small blocks $\mathcal{V}$ that contain $N_\mathcal{V}$ individual degrees of freedom with a discrete alphabet of $n$ symbols. Indeed:
\begin{equation}\label{RSMI_is_small}
	I_\Lambda(\mathcal{H}:\mathcal{E})\leq I(\mathcal{V}:\mathcal{E})=H(\mathcal{V})-H(\mathcal{V}|\mathcal{E})\leq H(\mathcal{V})\leq N_\mathcal{V}\log n,
\end{equation}
where, since $\Lambda$ compresses $\mathcal{V}$ into $\mathcal{H}$, the first step is the data-processing inequality, and the second inequality follows from the positive semi-definiteness of Shannon entropies. This is important, as MI estimation by variational lower-bounds can suffer from a bias-variance trade-off in the opposite regime,~\emph{i.e.}~when the MI is large.

An alternative expression for $I(X:Y)$ is in terms of the Kullback-Leibler (KL) divergence $D_{\rm KL}$:\footnote{The Kullback-Leibler divergence is a measure of distance (but formally not a metric) between the two probability distributions in its argument.}
\begin{equation}
	I(X:Y)=D_{\rm KL}\left[p(x,y)||p(x)p(y)\right],
\end{equation}
The Gibbs' inequality $D_{KL}(p||q)\geq 0$ predicates on a useful interpretation of MI: given $(X,Y)$ jointly distributed according to $p(x,y)$, $I(X:Y)$ measures the information lost when encoding $(X,Y)$ as a pair of independent random variables while they may not be so. This is $0$ if and only if $X$ and $Y$ are actually independent, \emph{i.e.}~$p(x,y)=p(x)p(y)$.

\subsection{Noise-contrastive lower-bound of mutual information}\label{lowerbound_derivation}

The results in the text were obtained with the noise-contrastive lower-bound of MI (which we refer to as InfoNCE). Here we introduce it and briefly motivate its form. We begin by casting the MI as follows:
\begin{equation*}
	I(X : Y) = \mathbb{E}_{p(x,y)}\left[\log \frac{p(x|y)}{p(x)}\right] = \mathbb{E}_{p(x,y)}\left[\log \frac{p(y|x)}{p(y)}\right].
\end{equation*}
To proceed, let the conditional probability distribution $q(x|y)$ be a variational \textit{ansatz} for the conditional probability distribution $p(x|y)$ appearing in the definition of MI above. Consider the KL divergence between $p(x|y)$ and $q(x|y)$:
\begin{align}
	D_{\rm{KL}}(p(x|y)||q(x|y))& = \mathbb{E}_{p(x|y)}\left[\log \frac{p(x|y)}{q(x|y)}\right]\nonumber\\
	&=\sum_x \frac{p(x,y)}{p(y)}\left(\log p(x|y) - \log q(x|y)\right),
\end{align}
where the joint and the marginal distributions are related by %the Bayes' rule
$p(x,y)=p(x|y)p(y)=p(y|x)p(x)$. The variational distribution $q(x|y)$ serves to approximate $p(x|y)$: the KL divergence vanishes for $q^*(x|y)=p(x|y)$.  Positive semi-definiteness of the KL divergence yields a lower-bound for $I(X:Y)$, known as the Barber-Agakov (BA) bound:\cite{NIPS2003_2410}
\begin{align}
	I(X:Y)\geq& \mathbb{E}_{p(x,y)}\left[\log \frac{q(x|y)}{p(x)}\right]\nonumber\\
	&=\mathbb{E}_{p(x,y)}\left[\log q(x|y)\right]+H(X)=:I_{\rm BA}(X:Y),
\end{align}
where $H(X)$ is the entropy of the random variable $X$. $I_{\rm BA}$ is a functional of the marginal distribution $q(x|y)$
\begin{equation*}
	I_{\rm BA}(X:Y)=I_{\rm BA}(X:Y)[q(x|y)].
\end{equation*}
Since $D_{\rm KL}=0$ if and only if $q(x|y)=p(x|y)$, the BA bound is tight only for the optimal \textit{ansatz} $I_{\rm BA}(X:Y)[q(x|y)=p(x|y)]=I(X:Y)$.

MI being the KL divergence between the joint and the product of marginal distributions motivates the idea of circumventing the exhaustive modelling of $p(x|y)$ by $q(x|y)$, focusing instead on the correlations between the variables $X$ and $Y$. Consider then a variational \textit{ansatz} $q(x|y)$ constrained into an energy-based family of functions:
\begin{equation}\label{EBA_ap}
	q(x|y):=\frac{p(x)}{Z(y)}e^{f(x,y)},
\end{equation}
with the partition function $Z(y):=\mathbb{E}_{p(x)}\left[e^{f(x,y)}\right]$.
By forcing $q(x|y)$ to this form, the complex correlations within the possibly high-dimensional data $X$ are stripped-out into the marginal distribution $p(x)$. The resulting lower-bounds are sensitive mainly to the variables' interdependency. In other words, maximising the lower-bound of MI is rephrased as a search for a function $f(x,y)$ modelling the relationships (or shared information) between $X$ and $Y$ very well, ignoring the surplus relationships \textit{within} $X$ or $Y$  (whose contribution to the MI are insignificant). As noted in Ref.~\onlinecite{poole2019variational}, this conceptual step underlies the leap in terms of practical efficiency gained in the corresponding energy-based MI lower-bounds \emph{e.g.}~by Nguyen, Wainwright and Jordan (NWJ)\cite{5605355} and tractable unnormalised BA (TUBA) bound.\cite{poole2019variational}. 

Even so, such so-called single-sample bounds are known to suffer from the large variance of the resulting MI estimators.\cite{poole2019variational, oord2018representation} An improved approach is to divide a single batch of MC samples for the pair of random variables $(X,Y)$ into minibatches of K-fold replicated random variables $(X_i,Y_i)_{i=1}^K$, and to derive the corresponding ``multi-sample" lower-bounds (note the confusing, but standard, dual usage of the term ``sample"). These are obtained by taking the average of the single-sample bounds introduced above, and address the issue of large variance by means of noise-contrastive estimation (NCE)\cite{899a65b4919f47c8a06d115df85dad11}, as first proposed in the context of MI estimation in Ref.~\onlinecite{oord2018representation}. 

More concretely, a multi-sample (or replica) bound estimates $I(X_1,Y)$, where $(X_1,Y)\sim p(x_1,y)$, 
given $K-1$ additional independent replicas for one of the random variables, say $X$ (drawn from the marginal distribution), which are gathered into a multidimensional variable: $X_{2:K}\sim \prod_{j=2}^K p(x_j)$%=:r^{K-1}(x_{2:K}).

Considering the average of $I_{\rm NWJ}(X_i:Y_i)$ over the $K$ replica random variables such that $(X_i, Y_i)\sim p(x_i,y_i)$, \emph{i.e.}~with each $Y_j$ playing the role of $Y$ in turn, leads to the InfoNCE lower-bound of MI:\cite{poole2019variational}
\begin{equation}\label{InfoNCE_end}
	I(X:Y)\geq I_{\rm NCE}(X:Y):=\langle I_{\rm NWJ}(X:Y) \rangle =\frac{1}{K} \mathbb{E}_{\prod_{k=1}^K p(x_k, y_k)}  \left[\sum_{j=1}^K \log \frac{e^{f(x_j,y_j)}}{\frac{1}{K}\sum_{i=1}^{K}e^{f(x_i,y_j)}}\right].
\end{equation}
%\begin{equation}\label{InfoNCE_end}
%I(X:Y)\geq I_{\rm NCE}(X:Y):=\langle I_{\rm NWJ}(X:Y) \rangle =\frac{1}{K} \mathbb{E}_{\prod_{k=1}^K p(x_k, y_k)}  \left[\sum_{j=1}^K \log \frac{e^{g(x_j,y_j)}}{\frac{1}{K}\sum_{i=1}^{K}e^{g(x_i,y_j)}}\right].
%\end{equation}
Maximisation of the InfoNCE lower-bound for RSMI is the method used in the main text, where maximisation is over the discriminator functions $f$. A key algorithmic idea here, originally introduced in Ref.~\onlinecite{belghazi2018mine}, is to parametrize them by neural networks $f_{\Theta}$ and instead optimize over the network parameters $\Theta$ using stochastic gradient descent.

\subsection{Deep neural network architecture for RSMI optimisation}
Our goal in using the lower-bounds derived above is to optimize the RSMI, \textit{i.e.}~MI between the random variable $\mathcal{H}$ representing the coarse-grained degrees of freedom in the block $\mathcal{V}$, and $\mathcal{E}$ representing the degrees of freedom in the environment. 

There exist several admissible multi-layer perceptron (MLP) architectures for the corresponding energy-based \textit{ansatz} $f\equiv f_\Theta(h,e)$ (see above), with a set of variational parameters $\Theta$. Here, we opt for a \textit{separable} form, such that: 
\begin{equation}
	f_\Theta(h,e)=v^{\rm T}(h)u(e), 
\end{equation}
where $v$ and $u$ are array-valued functions (here, neural networks, whose weights constitute $\Theta$) that depend only on hidden variables and the environment, respectively. The networks $v$ and $u$ independently map $\mathcal{H}$ and $\mathcal{E}$ to a so-called \textit{embedding space}. The advantage of this choice is the ability to construct the elements of the scores matrix, storing the values of the \textit{ansatz} $f_\Theta$ for all pairs of jointly and independently drawn samples, in $N$ passes of the MLP ($N$ passes for both $v$ and $u$ networks) for a sample dataset of size $N$. This is in contrast to the requirement of $N^2$ passes for all $N(N-1)$ independent samples and $N$ joint samples in a \textit{concatenated} architecture $f_\Theta(h,e)=f_\Theta([h,e])$, where the data for the pair of variables are concatenated before they are fed to the MLP.

\begin{table}[h]
	\centering
	\caption[Details of the MLP architecture for RSMI lower-bound.]{\textbf{Details of architecture for the RSMI-estimation module of RSMI-NE.} The multi-layer perceptron architecture expresses the variational \textit{ansatz} $f_\Theta(h,e)=v^{\rm T}(h)u(e)$ for the RSMI lower-bound. For each of $v$ and $u$, we use a single fully-connected hidden layer, which is a tensor of shape (hidden dimension=32, embed dimension=8). $f_\Theta(h,e)$ is obtained by taking the inner product along the embedding axis. The neurons are activated by the ReLU function. We use the InfoNCE lower-bound, hence the baseline function is the constant $e$.}
	\begin{tabular}{|c|c|c|c|c|c|}
		\hline
		lower-bound &\textit{ansatz} type &$\#$ layers & hidden dim. & embed dim. & activation function  \\
		\hline
		\hline
		InfoNCE & separable & 2 & 32 & 8  & ReLU  \\
		\hline
	\end{tabular}
	\label{concat_arch_parameters}
\end{table}

Tabulated in Tab.\ref{concat_arch_parameters} are the details of the network architecture for $f_\Theta(h,e)$ that we have used in this work. Nevertheless, we note that the results of the RSMI-NE are not very sensitive to a specific choice of these parameters. Specifically, we opted for two hidden layers each with 32 neurons fully-connected to the layer containing the $(\mathcal{H},\mathcal{E})$ data. The embedding dimension is 8. The neurons are activated by the rectified linear unit (ReLU) function (see,\textit{ e.g.}~Ref.~\onlinecite{Goodfellow-et-al-2016}).

\subsubsection{Gumbel-softmax reparametrisation trick for discretization of coarse-grained variables}
In the RSMI-NE architecture the coarse-grained variables $h$ are fed into the MI estimator. Since the MI value depends on what kind of distribution $h$ belongs to, we need to ensure this estimation step is not falsified by \emph{e.g.}~neglecting to force the output of the coarse-grainer into a discrete form, rather than a real number, if we decided $h$ to be Ising spins. The apparent problem is that discreteness of $h$ seems to spoil the differentiability of the whole setup. This is somewhat similar to the problem encountered in Variational Autoencoders (VAEs), which is solved there using the so-called \emph{reparametrization trick}, effectively allowing to only differentiate w.r.t.~to the parameters of the latent space probability distribution. This is the intuition behind the solution to the issue in RSMI-NE, which goes under the name of \emph{Gumbel-softmax reparametrisation trick}.\cite{NIPS2014_5449}

Let $h$ be a categorical random variable which can be in one of the states $\{i\}_{i=1}^N$ with the set of probabilities $\{\pi_i\}_{i=1}^N$. Given $\{g_i\}_{i=1}^N$, random variables drawn from the Gumbel distribution \cite{gumbel1954,NIPS2014_5449}, we define a vector-valued random variable utilizing the softmax function, whose j-th component takes the form:
\begin{equation}
	{\rm softmax}_{j,\epsilon}\left(\{g_i + \log \pi_i\}_{i=1}^N\right)=\frac{\exp\left[(\log \pi_j + g_j)/\epsilon\right]}{\sum_{i=1}^N\exp\left[(\log \pi_i + g_i)/\epsilon\right]},
\end{equation}
%\ref{eq:si_softmax}
where $\epsilon$ is the smearing parameter. For $\epsilon \rightarrow 0$ the softmax becomes the argmax function, mapping the argument vector $y=\{g_i + \log \pi_i\}_{i=1}^N$ into a $N$-component one-hot vector (one-hot encoding maps each of $N$ possible states $i$ of a discrete variable into a $N$-dimensional vector, with $1$ on $i$-th position, and zeros elsewhere) with some $k^*$-th entry taking the value $1$, thereby marking $y_{k^*}=\max y$. The resulting random variable is called a Gumbel-softmax random variable; it is only approximately (or pseudo-) discrete, for small enough $\epsilon$ (do not confuse with a discrete random variable defined by taking the maximum component of the softmax function). In practice though, the error coming from using a finite $\epsilon$ can be made comparable to machine precision. The samples from the Gumbel-softmax approximation of a certain categorical distribution $\{\pi_k\}_{k=1}^N$ are approximately one-hot vectors for small $\epsilon$. 

We anneal $\epsilon$ during the training, from $\epsilon_{\rm max}$ to $\epsilon_{\rm min}$, exponentially, with a decay exponent $r$.

\subsubsection{Network architecture for coarse-graining}\label{CG_network}
In Tab.~\ref{CG_arch} we tabulate the details of the architecture for the coarse-graining network we used for the 2D dimer model (and other examples in the companion paper Ref.~\onlinecite{rsmine}). We stack a single layer convolutional neural network (CNN) (generally with multiple kernels, corresponding to different components of $\mathcal{H}$) and the Gumbel-softmax reparametrisation layer to embed the components of $\mathcal H$ into (pseudo-) binary variables. While we determine the relaxation parameter $r$ by experimentation and fix it for all models, we tune the initial value of the Gumbel-softmax temperature $\epsilon_{\rm max}$ according to the total number of iterations during training. 

\begin{table}[H]
	\centering
	\caption{\textbf{Architecture details of the coarse-graining module of RSMI-NE for the 2-d interacting dimer model on a square lattice.}}
	\begin{tabular}{|c||c|c|}
		\hline
		$L_{\mathcal{V}}$   & 8  \\
		\hline
		$L_{\mathcal{B}}$  & $\{2,4,6,8\}$ \\
		\hline
		$L_{\mathcal{E}}$ & 4  \\
		\hline
		number of components of $\mathcal{H}$ & 2   \\
		\hline
		embedding of $\mathcal H$  &  binary \\
		\hline
		$(\epsilon_{\rm max}$, $\epsilon_{\rm min})$ for Gumbel-softmax (GS) & $(0.75,0.1)$ \\
		\hline
		GS annealing parameter $r$ & $5\times 10^{-3}$ \\
		\hline
	\end{tabular}
	\label{CG_arch}
\end{table}

Representing the coarse-graining map using a CNN enables interpreting the weights of the network to be directly understood in terms of a generalized Kadanoff block-spin construct for real-space RG.\cite{Kadanoff:1976vx} Nevertheless, the RSMI principle does not restrict the specific type of the variational ansatz for coarse-graining, and the CNN form is not imperative.

\subsection{Unsupervised learning scheme}\label{RSMI-net_training}
We now describe the (unsupervised) training, maximising our neural network \textit{ansatz} $I_\Lambda(\mathcal{H}:\mathcal{E})[f_\Theta]$. 
The inputs of the RSMI-NE can be \emph{e.g.}~the Monte Carlo (MC) samples from the desired model (as in the main text), but equally well experimentally measured data. As usual in numerical investigations, ensuring good quality sampling is important. Since we use the InfoNCE bound, the sampling is divided into mini-batches, each containing $K$ samples. We separate in each sample the visible patch $\mathcal{V}$ and its environment $\mathcal{E}$, dismissing a finite buffer separating them. Then a single mini-batch is denoted by the multi-dimensional random variable $(v_{1:K},e_{1:K})=(v_1,\cdots,v_K,e_1,\cdots,e_K)$.

Let $\Lambda^s$ and $\Theta^s$ be the network parameters for the coarse-graining, and the InfoNCE \textit{ansatz} $f$, respectively. At the outset we initialize these as tensors containing random numbers. At each step $s$ of training, the samples in the mini-batch are used to coarse-grain the samples $v_i$ into $h_i[\Lambda^s]$ and compute the scores matrix $F_{ij}(\Theta^s, \Lambda^s)=f(h_i[\Lambda^s], e_j;\Theta^s)$ for the InfoNCE \textit{ansatz} and the current values of the network parameters. Here $\Lambda^s$ and $\Theta^s$ denote value of the network parameters in $s$'th training step. In the scores matrix, the entries with $i=j$ denote the jointly drawn samples and the rest denote independently drawn samples for the coarse-grained degree of freedom and the environment.

The \textit{ansatz} for RSMI can be formulated in a way which fixes the allowed alphabet from which the coarse-grained degrees of freedom are drawn. In our implementation, this is done by choosing a layer $\tau$ to generate $\mathcal{H}$. For discrete-valued degrees of freedom we use the Gumbel-softmax layer with an annealing schedule as described above. More generally, one can specify the number of convolutional channels according to the symmetries of the system.

The InfoNCE prediction (that of $p(h,e)$ being equal to $p(h)p(e)$ or not) for the mini-batch is computed via the scores matrix as (compare Eq.~\ref{InfoNCE_end}):
\begin{equation}\label{NCE_RSMI_prediction}
	Q(h_{1:K},e_{1:K};\Theta^s, \Lambda^s) = \sum_{j=1}^K \frac{F_{jj}(\Theta^s, \Lambda^s) }{\sum_{ij=1}^{K}\exp{F_{ij}(\Theta^s, \Lambda^s)}}.
\end{equation}
Then 
\begin{equation*}
	\log Q(h_{1:K},e_{1:K};\Theta^s, \Lambda^s) + \log K
\end{equation*}
gives our single mini-batch estimate of RSMI.

The gradients of the mini-batch estimate of RSMI with respect to $\Lambda$ and $\Theta$ are then used for proposing the updated set of network parameters. More concretely, we use the adam optimizer\cite{kingma2014adam} to perform stochastic gradient-ascent. We have found that using the same learning rate for both parameter sets $\Lambda$ and $\Theta$ leads to efficient training.

We repeat this procedure over all mini-batches until all samples are fed to the network once. This constitutes one epoch of training. We repeat it for multiple epochs until convergence (see Ref.~\onlinecite{rsmine} for a more detailed discussion of convergence criteria), when we are left with an optimized coarse-graining filter represented by the final convolutional network parameters $\Lambda$, and an estimate of the RSMI given by a moving average of the time-series of mini-batch estimates. 

\section{Dependence on the buffer, block and system size}\label{section:scaling}
Here we provide a more detailed discussion of the length scales set by the linear sizes of the buffer $\mathcal{B}$ and the block $\mathcal{V}$, respectively denoted by $L_\mathcal{B}$ and $L_\mathcal{V}$, and their physical interpretation as well as practical importance.

The nature of the scales $L_B$ and $L_\mathcal{V}$ is different. $L_B$ sets the distinction between long- and short-range contributions to RSMI. Naively, one may assume $L_B \rightarrow \infty$ is necessary, but this is not entirely true. To determine the functional form of the operator in terms of local degrees of freedom, the scale $L_B$ needs only to be sufficiently large, so that correlations due to the \emph{second-most} relevant operator decayed to the point of being indistinguishable from statistical noise always present in the sampling, while those of the most relevant one are still discernable (assuming the leading operator is not degenerate). For a given total system size, further increasing $L_B$ cannot change the answer, as the most relevant operator has already ``won", and is the sole contributor to RSMI. In fact, with finite sampling, it would make the problem computationally harder, or impossible, as the leading correlations would be drowned by the noise. Moreover, even when $L_B$ is not large enough to perfectly distinguish leading from subleading contributions, \emph{both of them} can be extracted using the \emph{filter ensemble} analysis descrived in the companion manuscript Ref.~\onlinecite{rsmine}. Thus, in practice, already small buffers of a dozen sites are sufficient to find the correct operator content (but not necessarily their correct dimensions yet, see below). In use, one should systematically scan (increase) the size of $L_B$.

The block $V$, on the other hand, is the support (in the mathematical sense) of the relevant operator/order parameter. Assuming this operator is \emph{local}, in the first instance $V$ needs only to be sufficiently large so that the appropriate function of the original degrees of freedom can be constructed. For example, the ``plaquette" operators in the dimer model require a block twice smaller than what we used, and the plaquette pattern is repeated in the block. This effectively corresponds to computing \emph{block-averages} of the operator values, which produces quasi-continuous outputs even in completely discrete systems. The degree of smearing is controlled by $L_{\mathcal{V}}$. This is discussed in detail in Sec.III.D.1 and Sec.III.D.2 of the companion Ref.\onlinecite{rsmine}, where we show it allows to recover continuous emergent symmetries.
Note also that for \emph{e.g.}~experimental measurements of a quantity performed on a grid of an arbitrary spatial resolution, the relevant operator/order parameter may require a large support in such arbitrary units. As for $L_B$, one should then scan the block size, or set it based on prior knowledge. In typical cases, the size $V$ is very small though.

Though $L_B$ and $L_\mathcal{V}$ need not be large to extract the operators, the above \emph{does not imply} that larger system sizes are superfluous. In smaller systems, certain operators may appear more relevant than others in parameter regimes (\emph{e.g.}~above a phase transition) where they would be already subleading, had the system been closer to the thermodynamic limit (see plaquette filters in Fig.~2c in the main text). The correctness and accuracy of determination of their scaling dimension from the correlator will be improved with the system size (and the number of samples). We emphasize, however, that RSMI-NE discovers the parametrized form, in terms of the original local degrees of freedom, of both the leading and subleading operators. This allows to ``export" them to analyse their properties with other methods (even analytically, perhaps). We also expect that standard MC scaling analysis to simplify, as we already have \emph{the correct operators} and only need to compute their correlators. To wit, we fit for the charge operator $\mathcal{O}_1$ at $T\rightarrow\infty$ a dimension of 1.00037, in excellent agreement with the predicted value of  $1.0$, already at system size of $64\times64$.
Thus, the method has very favorable properties, extracting parametrized leading and subleading operators on the lattice even from modest system sizes, but large systems and more samples allow to more accurately determine their properties, particularly the scaling dimensions.

\section{Details of operator extraction in the interacting dimer model}\label{section:dimer_further}

\subsection{Dimer model: Height-field mapping of dimers on 2-d square lattice}
\begin{figure}[h!]
	\centering
	\includegraphics[width=1\linewidth]{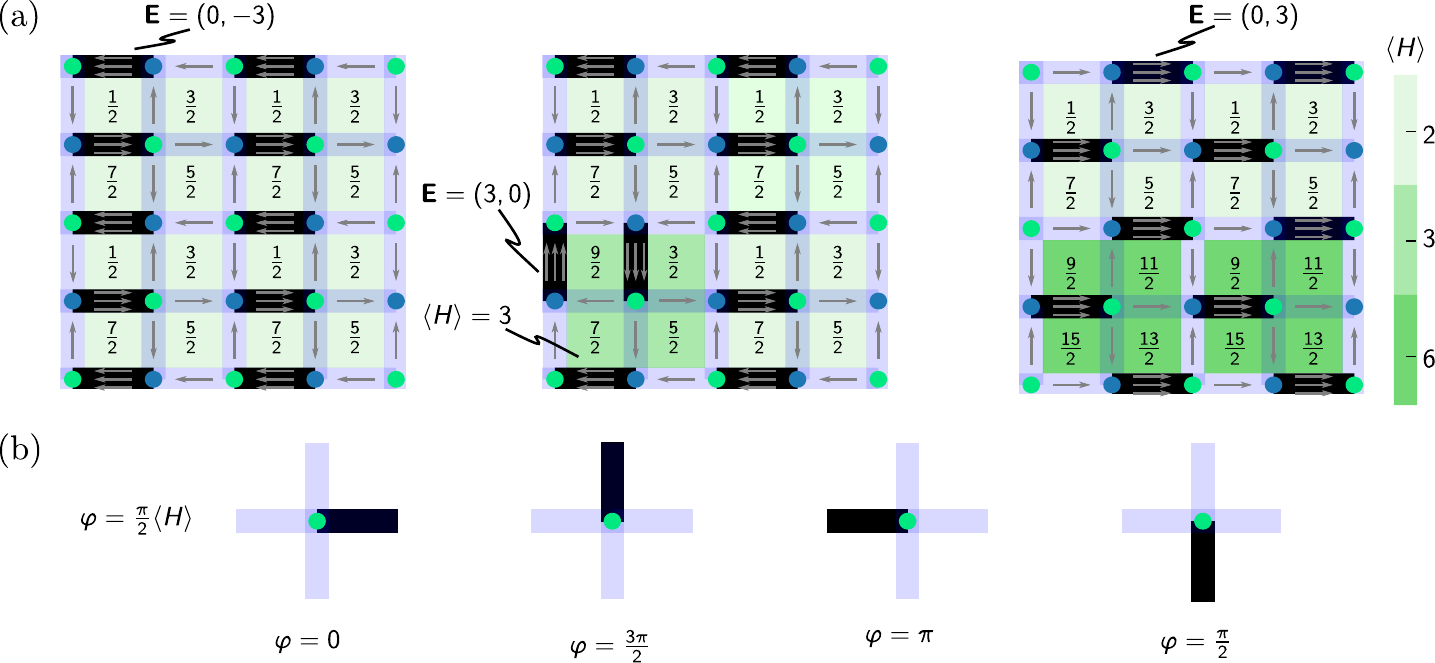}
	\caption{{\bf Height field mapping of the dimer model on a square lattice.} {\bf a} The height fields are defined on the plaquettes, and the electric fields related to their gradients are shown by the arrows on the bonds (see explanation in the text). The color-map shows the average height $\langle H(\mathbf{r}_i)\rangle$ over the four plaquettes surrounding an even-sublattice site $i$ (denoted by green dots). Left: for a columnar ground-state configuration the height profile $\langle H(\mathbf{r}) \rangle$ is \emph{uniform}. Correspondingly, the average electric field vanishes over the whole configuration.  Middle: upon flipping a pair of dimers around a plaquette, $\langle H(\mathbf{r}) \rangle$ changes by one. Right: staggered configurations have a net electric field over the whole configuration (observe the orientation of the arrows). Correspondingly the height field has a \emph{tilted} profile. {\bf b} $\varphi=\frac{\pi}{2}\langle H \rangle$ is interpreted as an angle and is directly related to the orientation of the dimer at the given site.}
	\label{fig:heightfield_map}
\end{figure}

Here we describe the classic mapping of the dimer models to height fields, as per \emph{e.g.}~Ref.~\onlinecite{fradkin_2013}. Since the 2-d square lattice is bipartite, we can map each dimer configuration $C(\mathbf{r}_i)$ onto a unique height profile $H(\mathbf{r}_i)$ living on the plaquettes of the lattice. Without loss of generality, we start by assigning a reference height $H(\mathbf{r}_0)=1/2$ to the plaquette at $i=0$. The heights of the neighboring plaquettes are then determined by winding clockwise (counter-clockwise) around a vertex of the even (odd) sublattice, and changing the height by $-3$ if a dimer is crossed on the link connecting the two plaquettes, or changing by $+1$ otherwise, see Fig.~\ref{fig:heightfield_map}. The fact that each site in the fully packed dimer model can have a single dimer with only one of the four orientations attached to it is reflected in the height profile as an invariance under a uniform shift by 4 of the entire profile: $H(\mathbf{r})\equiv H(\mathbf{r})+4$. This makes it natural to define the $2\pi$ periodic rescaled field $\varphi(\mathbf{r}_i)=\frac{\pi}{2}H(\mathbf{r}_i)$, which can be interpreted as the orientation angle of the dimer connected to a given site. As shown in Fig.~\ref{fig:heightfield_map}, we can assign to each site in the odd sublattice the average height of the four plaquettes that surround it. This assignment takes four distinct values $\left\{0, \frac{\pi}{2},\pi, \frac{3\pi}{2}\right\}$, respectively corresponding to $\{\text{right, up, left, down}\}$ oriented dimers connected to the site.

The mapping can also be realized in terms of the electric fields on the links, see Fig.~\ref{fig:heightfield_map}. Imagine giving each dimer a fixed orientation from even to odd sites. Then the above mapping corresponds to assigning 3 units of electric field pointing along the orientation of the dimer (\emph{i.e.}~flowing from the even to the odd sublattice site), and 1 unit of electric field, flowing out along the three remaining unoccupied links connected to each odd site. It follows that for fully packed dimers these electric fields locally conserve the flux, \emph{i.e.} have a vanishing lattice divergence $\nabla \cdot \mathbf{E}=0$, which is the Gauss' law. 
Observe that the height can be treated as a scalar potential for this field: $\nabla \varphi = \mathbf{E}$ if we identify $\mathbf{E}=(E_{|}, E_{-})$, that is: the gradient of the height field in x-direction is the y-component of the electric field, and \emph{vice versa} (see Fig.~\ref{fig:heightfield_map}). This height mapping serves as a basis for deriving the continuum effective field theory, ultimately giving the sine-Gordon action in the main text.\cite{PhysRevB.76.134514,fradkin_2013}

\subsection{Order parameters from RSMI-optimal coarse-graining filters}
As mentioned in the main text, any two-component vector of filters with the components drawn from among the columnar (C) and the two plaquette filters (P1, P2) defines a coarse-graining which deterministically encodes the four columnar ground-states without any ambiguity, as shown in Tab.1 in the main text. As all these filters recover two bits of information, they occur degenerately in the long-range ordered phase (LRO). 

In fact, the filters can be interpreted as components of an order parameter, which distinguish the four distinct sectors of the phase space arising due to the spontaneous breaking of the $C_4$ symmetry. More concretely, one can define a global two component \emph{plaquette} order parameter $(D_1, D_2)$:
\begin{equation}
	D_i:=\mathbb{E}\left[\frac{1}{N_{\mathcal{V}}}\sum_{k} \tau \circ \left(\Lambda_i\cdot \mathcal{V}_k\right)\right],
\end{equation}
using \emph{e.g.}~$\Lambda_1 = \Lambda_{\rm C}$ or $\Lambda_1 = \Lambda_{\rm P1}$, and $\Lambda_2 = \Lambda_{\rm P2}$, where the sum inside the expectation is taken over all disjoint blocks $\mathcal{V}_k$ on the lattice, and $N_{\mathcal{V}}$ is the number of such blocks. The norm of this order parameter is plotted as order parameter $P$ in Fig.~2.f in the main text. The columnar filter $\Lambda_{\rm C}$ alone defines the dimer orientational symmetry breaking (DSB) order  parameter, exactly equivalent to the one introduced by Alet \emph{et al.}\cite{PhysRevE.74.041124} (also plotted in Fig.~2.f in the main text):
\begin{equation}
	{\rm DSB}:=\mathbb{E}\left[\sum_k \tau \circ \left( \Lambda_{\rm C}\cdot \mathcal{V}_k\right)\right].
\end{equation}

Although both columnar and plaquette filters signal the columnar order with a non-vanishing expectation value, their behavior differs above the BKT transition temperature. As shown in Fig.2.f in the main text, while DSB decays quickly to $0$ for $T>T_{\ssm BKT}$ even for small lattices the plaquette ordered parameter decays much more slowly. This distinction is readily understood. The lowest-lying excitations above the columnar ground states are plaquette flips, which cost two units of energy. However, the mappings defined by the $\Lambda_{\rm P1/P2}$ filters are invariant under subsets of plaquette flips in the configuration, and therefore the expectation  stays non-zero for finite-size systems at $T>T_{\ssm BKT}$ (in an infinite system the plaquette order parameter also decays to zero).

\subsection{Discovering the relevant operators}
\paragraph{Plaquette/columnar filters, and  the electric charge operators.}
\begin{table}[H]
	\centering
	\caption{\textbf{Scaling dimensions of the electric charge operators at the BKT transition, and at the free dimer point.}}
	\label{tab:dimer_op_exponents}
	\begin{tabular}{|c||m{1cm} m{1cm}|}
		\hline
		electric charge operator &$ \mathcal{O}_1$ & $\mathcal{O}_2$ \\
		\hline
		\hline
		$T\to \infty$ & $d_1=1$ & $d_2=4$ \\%[2ex] 
		$T=T_{\ssm BKT}$ & $d_1=\frac{1}{8}$ & $d_2=\frac{1}{2}$ \\%[2 ex] 
		\hline
	\end{tabular}
\end{table}

For $T\geq T_{\ssm BKT}$, the temperature-dependent RG scaling dimensions of the $\mathcal{O}_{n=1,2}$ operators, given in Ref.~\onlinecite{PhysRevB.76.134514}, are proportional to $n^2$. We list the values they take at the BKT transition, and in the limit of infinite $T$ in Tab.~\ref{tab:dimer_op_exponents}. This field-theoretic result underlies the evolution of the RSMI-optimal filters with $T$. The orientational symmetry breaking and plaquette operators are degenerate in the long-range ordered phase; in the critical phase, however, the columnar order parameter corresponding to $n=2$ has the higher scaling dimension, therefore its correlations decay faster, and this is why we obtain the plaquette filters but not the columnar ones beyond the BKT point. We thus verify for the case of interacting fully-packed dimers that the RSMI-NE indeed finds the operators with the lowest-scaling dimension, \emph{i.e.}~the RG relevant operators.

\begin{figure}[h!]
	\centering
	\includegraphics[width=0.9\linewidth]{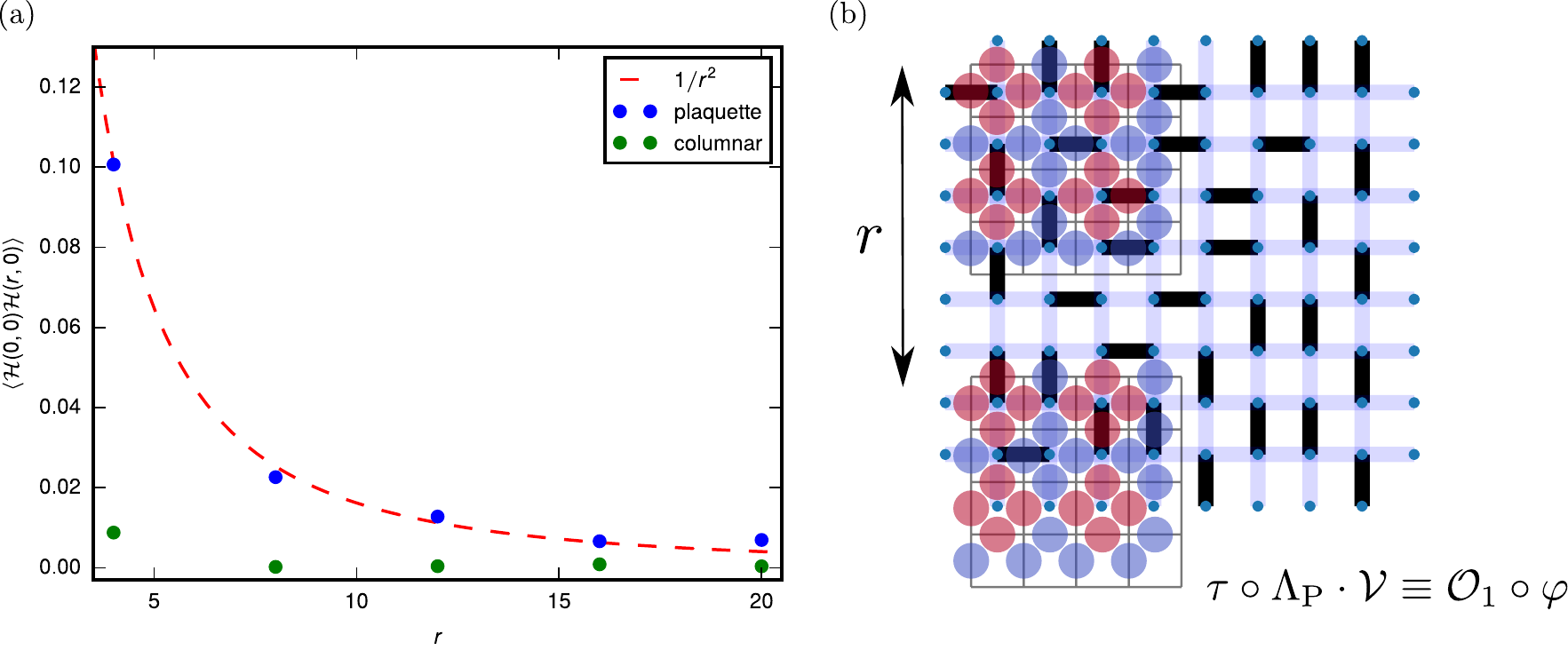}
	\caption{{\bf Using RSMI filters to compute correlation functions and scaling dimensions:} The plaquette and the columnar coarse-graining filters correspond to the discretized $\mathcal{O}_1$ and $\mathcal{O}_2$ operators, with scaling dimensions $d_1=1, d_2=4$ at $T\to \infty$. (a) The correlators obtained using these filters confirm this picture. We fit a power-law decay to the plaquette correlator (on a $128\times 128$ dimer system, using size $4\times 4$ filters), obtaining a best-fit of $2.00074$, in excellent agreement with the field theoretic result $2d_1=2$, thus extracting the scaling dimension. The columnar correlator is consistent with the theoretical prediction of $r^{-8}$, but the fast decay would necessitate larger sample size for an accurate fitting of the exponent. (b) More precisely, the correlation function for $\mathcal{O}_{1,2}$ on the height field $\varphi$ is directly computed from the MC samples as the correlator of the trained neural networks parametrizing the coarse-graining filters $\Lambda_{\rm P, C}$, evaluated at spatial separations $r$ on the dimer configurations $C$.}
	\label{fig:plaqcolcorrelators}
\end{figure}

We emphasize that the extracted RSMI filters can be used exactly as one usually uses operators. For instance, their correlation functions can be evaluated numerically from the samples, to recover (fit) their scaling dimensions. Though above we gave an explicit mapping between the filters and the scaling operators of the dimer model, so it is superfluous in this case, in more complex scenarios, with systems not fully understood, this would be necessary. As an illustration, in Fig.~\ref{fig:plaqcolcorrelators} we evaluate and plot the correlation function (on a $128\times 128$ dimer system, utilizing size $4\times 4$ filters) of the plaquette operators in the limit $T\rightarrow\infty$, and verify accurately it does indeed decay as $r^{-2}$, \emph{i.e.}~the we extract the scaling dimension $d_1 = 1$ for the $\mathcal{O}_1$ operator, a predicted.\cite{PhysRevB.76.134514} The decay of correlation function of the columnar filters is much faster, consistent with the prediction of $r^{-8}$ (we do not fit the exponent, as the fast decay requires better sample statistics). Thus, RSMI-NE allows to gain a formal understanding of the long-distance properties of the system, unsupervised and without inputting any prior knowledge.

\paragraph{Staggered filters, electric fields.}
The local action of pristine staggered filters which are RSMI-optimal at $T\to \infty$ has the form
\begin{align}\label{stag_filters}
	\Lambda_{\rm S+}\cdot \mathcal{V}(\mathbf{r}) &= (-1)^{x+y}N_{|}(\mathbf{r}),\\
	\Lambda_{\rm S-}\cdot \mathcal{V}(\mathbf{r}) &= (-1)^{x+y+1}N_{-}(\mathbf{r}).
\end{align}
Note that here $\left(\Lambda_{\rm S+}:=\frac{\Lambda_{\rm S1}+\Lambda_{\rm S2}}{\sqrt{2}},\, \Lambda_{\rm S-}:=\frac{\Lambda_{\rm S1}-\Lambda_{\rm S2}}{\sqrt{2}}\right)$ corresponds to the orthogonal basis $(E_x,E_y)$ for the electrical field.\footnote{S1 and S2 refers to the pristine staggered filters introduced in the main text.} On the other hand, any unitary rotation $R(\theta)(\Lambda_{\rm S+}, \Lambda_{\rm S-})\sim \left(\cos(\theta) E_x - \sin(\theta )E_y, \sin(\theta) E_x + \cos(\theta) E_y\right)$ recovers the same amount of RSMI, indicating the emergent $U(1)$ symmetry of the free dimer model. For the interested reader, we point Ref.~\onlinecite{rsmine} for more details.

Ref.~\onlinecite{PhysRevB.76.134514} gives an expansion of the dimer densities on the lattice in terms of the observables of the height-field continuum model:
\begin{align}
	N_{|}(\mathbf{r})=\frac{1}{4}+\frac{(-1)^{x+y+1}}{2\pi}\partial_{x}\varphi(\mathbf{r})+(-1)^y \sin \varphi(\mathbf{r})&,\\
	N_{-}(\mathbf{r})=\frac{1}{4}+\frac{(-1)^{x+y}}{2\pi}\partial_{y}\varphi(\mathbf{r})+(-1)^x \cos \varphi(\mathbf{r})&.
\end{align}
Here, $1/4$ is the average dimer density (due to the $C_4$ symmetry). The terms $(-1)^y\sin\varphi$ and $(-1)^x\cos\varphi$, respectively, select the vertical and horizontal dimers connected to point $\mathbf{r}$ (\emph{cf.} Fig.~\ref{fig:heightfield_map}.b).

Substituting these in Eq.~\ref{stag_filters} and averaging over the degrees of freedom in the block $\mathcal{V}$ to which the filter is applied, we obtain the coarse-grained degrees of freedom:
\begin{align}
	\mathcal{H}_1\sim\sum_{\mathbf{r}\in \mathcal{V}} \Lambda_{\rm S+}\cdot \mathcal{V}(\mathbf{r}) &= \sum_{\mathbf{r}\in \mathcal{V}}\left[\frac{(-1)^{x+y}}{4} + \frac{\partial_x \varphi(\mathbf{r})}{2\pi} + (-1)^x \sin \varphi(\mathbf{r})\right] = \sum_{\mathbf{r}\in \mathcal{V}}\left[\frac{\partial_x \varphi(\mathbf{r})}{2\pi} + (-1)^x \sin \varphi(\mathbf{r})\right],\\
	\mathcal{H}_2\sim\sum_{\mathbf{r}\in \mathcal{V}} \Lambda_{\rm S-}\cdot \mathcal{V}(\mathbf{r}) &= \sum_{\mathbf{r}\in \mathcal{V}}\left[\frac{(-1)^{x+y}}{4} + \frac{\partial_y \varphi(\mathbf{r})}{2\pi} + (-1)^y \cos\varphi(\mathbf{r})\right] = \sum_{\mathbf{r}\in \mathcal{V}}\left[\frac{\partial_y \varphi(\mathbf{r})}{2\pi} + (-1)^y \cos\varphi(\mathbf{r})\right],
\end{align}
where the $\sim$ symbol is used to denote that the coarse-grained variable $\mathcal{H}$ is the result of applying the binary mapping $\tau$ to the right hand side. We see that the first term in the summand vanishes since there are an even number of links in the region $\mathcal{V}$. The last term probes the tilt of the height configuration either in the $x$ or in the $y$ direction. 
This is because the term $(-1)^x \sin\varphi(\mathbf{r})$ has a non-zero average only if the region $\mathcal{V}$ contains a vertical dimer with a missing parallel neighbor, in which case the height profile has a slope in the $x$ direction, while it vanishes for a uniform height field (similarly for the other component in the $y$ direction). Equivalently, only not-constant-in-x function $\varphi(\mathbf{r})$ survives averaging with $-1^x$ over the block, and so under the average we may replace $\sin \varphi(\mathbf{r})$ with $\partial_x \varphi(\mathbf{r})$. Hence, in total:
\begin{equation}
	\mathcal{H} \propto \tau\circ \nabla \left\langle \varphi(\mathbf{r}) \right\rangle_{\mathbf{r}\in\mathcal V},
\end{equation}
where the average is over the sites in the block and $\tau$ maps the gradient into binary values. In other words, the staggered filters extract the coarse-grained gradients of the height field, \emph{i.e.}~they correspond to the electric fields described in Fig.~\ref{fig:heightfield_map}.a.

%TODO: Refer the reader to the PRE for the extensive discussion of the emergent $U(1)$ symmetry of the electrical fields.

\smallskip
\noindent
{\bf Acknowledgements} This work was supported by a grant from the Swiss National Supercomputing Centre (CSCS) under project ID eth5b. 

\bibliographystyle{naturemag}
\bibliography{ref}